\documentclass[final,5p,times,twocolumn]{elsarticle}

\usepackage{lineno,hyperref}
\modulolinenumbers[5]

\journal{Nuclear Instruments $\&$ Methods in Physics Research, Section A}

\begin{document}

\setlength\parindent{0pt}

\begin{frontmatter}

\title{Radiation length imaging with high-resolution telescopes}

\author[mymainaddress]{U. Stolzenberg\corref{mycorrespondingauthor}}
\cortext[mycorrespondingauthor]{Corresponding author}
\ead{ulf.stolzenberg@phys.uni-goettingen.de}
\author[mymainaddress]{A. Frey}
\author[mymainaddress]{B. Schwenker}
\author[mymainaddress]{P. Wieduwilt}
\author[mysecondaryaddress]{C. Marinas}
\author[mysecondaryaddress]{ F. L\"utticke}

\address[mymainaddress]{II. Physikalisches Institut, Universit\"at G\"ottingen, Friedrich-Hund-Platz 1, 37077 G\"ottingen, Germany}
\address[mysecondaryaddress]{Physikalisches Institut, Universit\"at Bonn, Nu{\ss}allee 12, 53115 Bonn, Germany}

\begin{abstract}
The construction of low mass vertex detectors with a high level of system integration is of great interest for next generation collider experiments. Radiation length images with a sufficient spatial resolution can be used to measure and disentangle complex radiation length $X$/$X_0$ profiles and contribute to the understanding of vertex detector systems. Test beam experiments with multi GeV particle beams and high-resolution tracking telescopes provide an opportunity to obtain precise 2D images of the radiation length of thin planar objects. At the heart of the $X$/$X_0$ imaging is a spatially resolved measurement of the scattering angles of particles traversing the object under study. The main challenges are the alignment of the reference telescope and the calibration of its angular resolution.\\
 In order to demonstrate the capabilities of $X$/$X_0$ imaging, a test beam experiment has been conducted. The devices under test were two mechanical prototype modules of the Belle II vertex detector. A data sample of 100 million tracks at $4\, \mathrm{GeV}$ has been collected, which is sufficient to resolve complex material profiles on the $30\,\mu$m scale.
\end{abstract}


\end{frontmatter}


\section{Introduction}
\label{sec:introduction}

Physics requirements limit the mass of vertex and tracking detectors for next generation collider experiments \cite{Cooper:2014uda}. To meet the conflicting goals of minimizing detector mass and also incorporating cooling, power distribution and readout electronics an integrated system design approach is needed \cite{Demarteau:2014pka}. The precision of the measured vertex and track parameters is degraded by multiple scattering effects. The magnitude of multiple scattering depends on the local material budget measured in units of the material constant $X_0$. It is therefore crucial to carefully limit and account for every material contribution in the detector acceptance region.\\
In order to study the material composition of planar objects a method has been developed to measure high-resolution 2D images of the radiation length $X$/$X_0$. The radiation length can be extracted from the multiple scattering deflections by using an appropriate theoretical model. The good spatial resolution of the method opens a way to measure the overall scattering effect in small areas. It can consequently be used to map fully passive materials very precisely. This provides an opportunity to disentangle material distributions from various different components like ASICs, bump bonds and support structures.

\section{Method}
\label{sec:method}

For the radiation length measurements the planar object is centered in a high-resolution tracking telescope. A detailed description of the data processing including clustering, alignment and tracking can be found in \cite{Schwenker14}. The basic idea is to reconstruct multiple scattering angles from charged particle tracks.  A pair of Kalman filters can be used to compute the track state and covariance matrix before the scattering $\eta^\mathrm{in}$, $C^\mathrm{in}$ from the upstream hits and the state and covariance matrix after the scattering $\eta^\mathrm{out}$, $C^\mathrm{out}$ from the downstream hits. The track states are extrapolated to the scattering plane of the object ($u$,$v$,$w$)  and defined as 
\begin{eqnarray}
\label{eq:trackstate}
\eta^\mathrm{in}&=&\left(\left({\mathrm{d}u\mathrm{/}\mathrm{d}w}\right)^\mathrm{in},\left({\mathrm{d}v\mathrm{/}\mathrm{d}w}\right)^\mathrm{in},u^\mathrm{in},v^\mathrm{in}\right) \\ \eta^\mathrm{out}&=&\left(\left({\mathrm{d}u\mathrm{/}\mathrm{d}w}\right)^\mathrm{out},\left({\mathrm{d}v\mathrm{/}\mathrm{d}w}\right)^\mathrm{out},u^\mathrm{out},v^\mathrm{out }\right) \nonumber \quad .
\end{eqnarray}
While the measurement results are obtained from the optimal weighted mean, for clarity of this text the following formula for the calculation of the intersection point is used, which neglects small correlation effects:
\begin{eqnarray}
\label{eq:intersection}
u&=&\left(u^\mathrm{in}+u^\mathrm{out}\right)\mathrm{/}2\\
v&=&\left(v^\mathrm{in}+v^\mathrm{out}\right)\mathrm{/}2 \nonumber
\end{eqnarray}
The two projected scattering angles $\vartheta_i$ ($i$=$u,v$) can be calculated from the difference of track slopes:
\begin{eqnarray}
\label{eq:angle}
\vartheta_u=\left({\mathrm{d}u\mathrm{/}\mathrm{d}w}\right)^\mathrm{out}-\left({\mathrm{d}u\mathrm{/}\mathrm{d}w}\right)^\mathrm{in} \quad \quad \left(\vartheta_v \, \mathrm{analogous }\right) 
\end{eqnarray}
In many cases, the correlation between the estimated scattering angles is very small and can be neglected. The angular resolution can then be computed from the following formula:
\begin{eqnarray}
\label{eq:resolution}
\sigma_\mathrm{reso}=\sqrt{C_{\frac{\mathrm{d}u}{\mathrm{d}w}\frac{\mathrm{d}u}{\mathrm{d}w}}^\mathrm{in} + C_{\frac{\mathrm{d}u}{\mathrm{d}w}\frac{\mathrm{d}u}{\mathrm{d}w}}^\mathrm{out}} \quad \quad \left(v \, \,\mathrm{component} \, \, \mathrm{analogous }\right) 
\end{eqnarray}
 This radiation length estimation approach is based on a $X$/$X_0$ estimation technique, which was introduced in \cite{Nadler2011246}. No hit from the scattering planar object is used, therefore the material distribution of any thin object can be analysed. Furthermore due to the track extrapolation from the upstream and downstream side no assumption of the material distribution of the planar object is required. \\
The object plane is divided into small pixel areas and the tracks are sorted into different pixels according to their intersection point (see eq. \ref{eq:intersection}). In order to compute the radiation length value of one pixel of the image the measured angle distribution of this pixel is fitted with the function: 
\begin{eqnarray}
\label{eq:fitfunction}
f_\mathrm{reco}\left(\vartheta\right)=\mathrm{f}_\mathrm{MSC}\otimes \frac{1}{\left(\lambda\sigma_\mathrm{reso}\right)\sqrt{2\pi}}\exp{\left(\frac{-\vartheta^2}{2 \left(\lambda\sigma_\mathrm{reso}\right)^2}\right)}
\end{eqnarray}
The finite telescope angular resolution $\sigma_\mathrm{reso}$ introduces a gaussian error, which broadens the multiple scattering angle distribution $\mathrm{f}_\mathrm{MSC}$. Hence, the fit function is a convolution between the multiple scattering distribution $\mathrm{f}_\mathrm{MSC}$ and the gaussian error function. The correction factor $\lambda$ is used to tune the angular resolution for a specific experimental setup to $\lambda\cdot\sigma_\mathrm{reso}$ (see section \ref{sec:setupandcalibration}). There are different models to describe multiple scattering distributions $\mathrm{f}_\mathrm{MSC}$. Two of the most prominent ones were introduced by Moliere \cite{Moliere1948} and Highland \cite{HIGHLAND1975497}. A recent review of multiple scattering models can be found in \cite{Frühwirth2001369}. Independently of the model, the width of the angle distribution is determined by the particle momentum $p$, the thickness $X$ of the material and material properties, which can be reduced to $X_0$. The width of the angular distribution also depends on the mass of the particle, which can be neglected only for light particles or for high momenta.

\section{Experimental setup and calibration}
\label{sec:setupandcalibration}

\begin{figure}[h!]
 	\centering
  	\includegraphics[width= 0.790\linewidth]{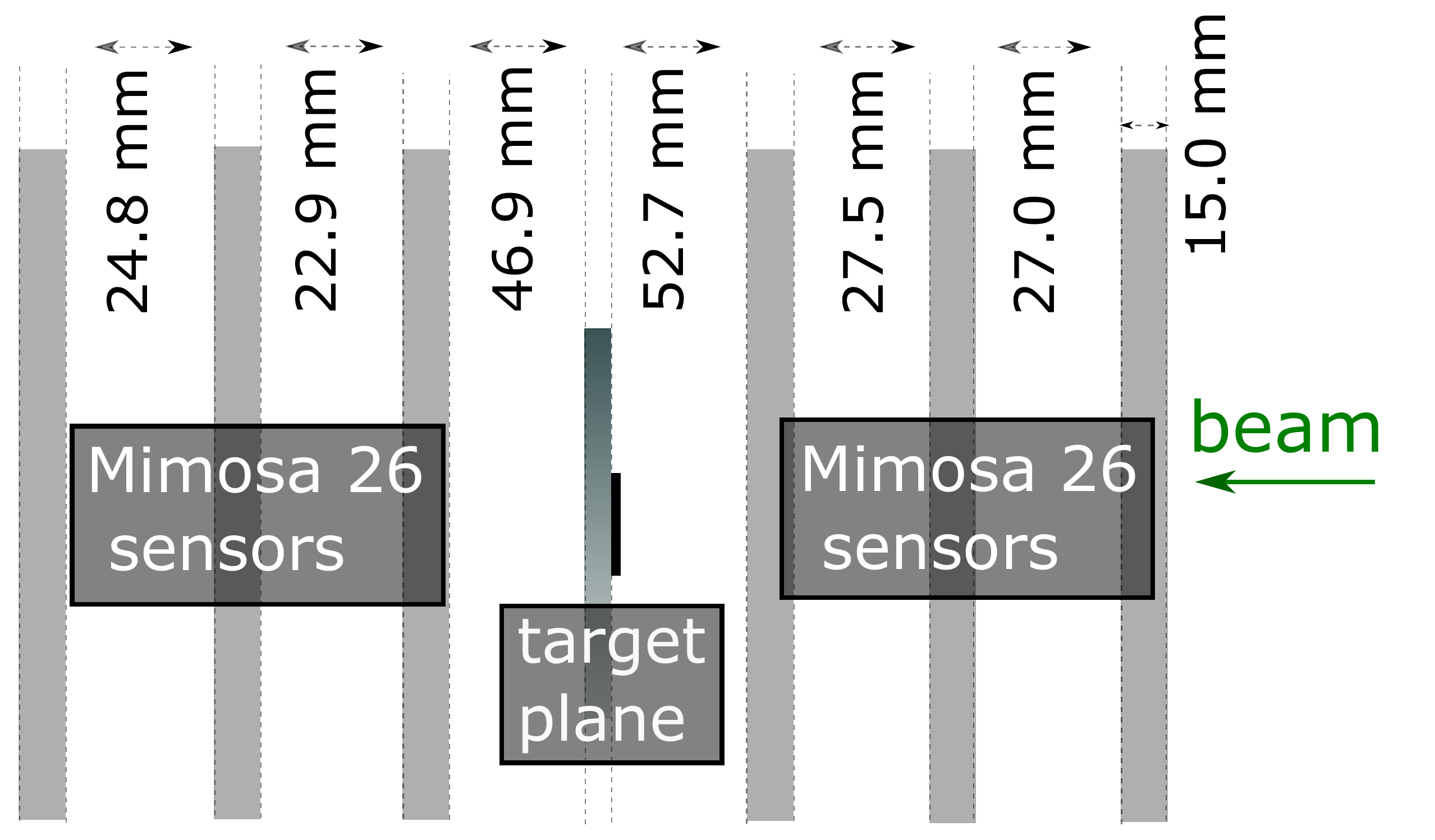}
 	\caption{{Experimental setup for $X$/$X_0$ imaging. The $X$/$X_0$ target is installed between the two arms of a high-resolution reference telescope and is supported perpendicular to the beam axis.}}
 	 \label{fig:Setup_PXD}
\end{figure}

For all the following measurements an AIDA telescope with 6 Mimosa26 (M26) sensors \cite{Rubinskiy2012923} was used. The zero suppression cut of the M26 sensors was tuned in such a way that an equal number of two pixel clusters and one pixel clusters for particle signals is achieved. This gives a hit coordinate resolution for M26 clusters of $3\,\mu$m with uncertainties $\sigma_\mathrm{M26}$ in the order of roughly $10\,\%$ \cite{Schwenker14}. \\
Due to the increased amount of multiple scattering for low energy particles, the method works best with a monoenergetic multi GeV particle beam like for example provided in the DESY test beam facilities. In this case a particle beam of $4\,$GeV electrons was employed. In preparation for the actual experiment toy studies were conducted to determine the resolution of the reference telescope at the target plane position depending on gaps between the telescope sensors. The selected telescope geometry (see figure \ref{fig:Setup_PXD}) is a compromise between a good spatial resolution of reconstructed tracks, favoring small spacings between the telescope planes, and a good angular resolution, which improves with the gap size. The spatial resolution of $4\,\mu$m is sufficient to resolve even small structures like bump bonds. The angular resolution of $134\,\mu$rad on the other hand is good enough to measure angles of 70 $\mu$rad, which correspond to the Highland multiple scattering angle width $\sigma_\mathrm{HL}$ for a scatterer consisting of $75\,\mu$m of silicon at a beam energy of $4\,$GeV. 

\begin{figure}[h!]
 	\centering
  	\includegraphics[width= 0.94\linewidth]{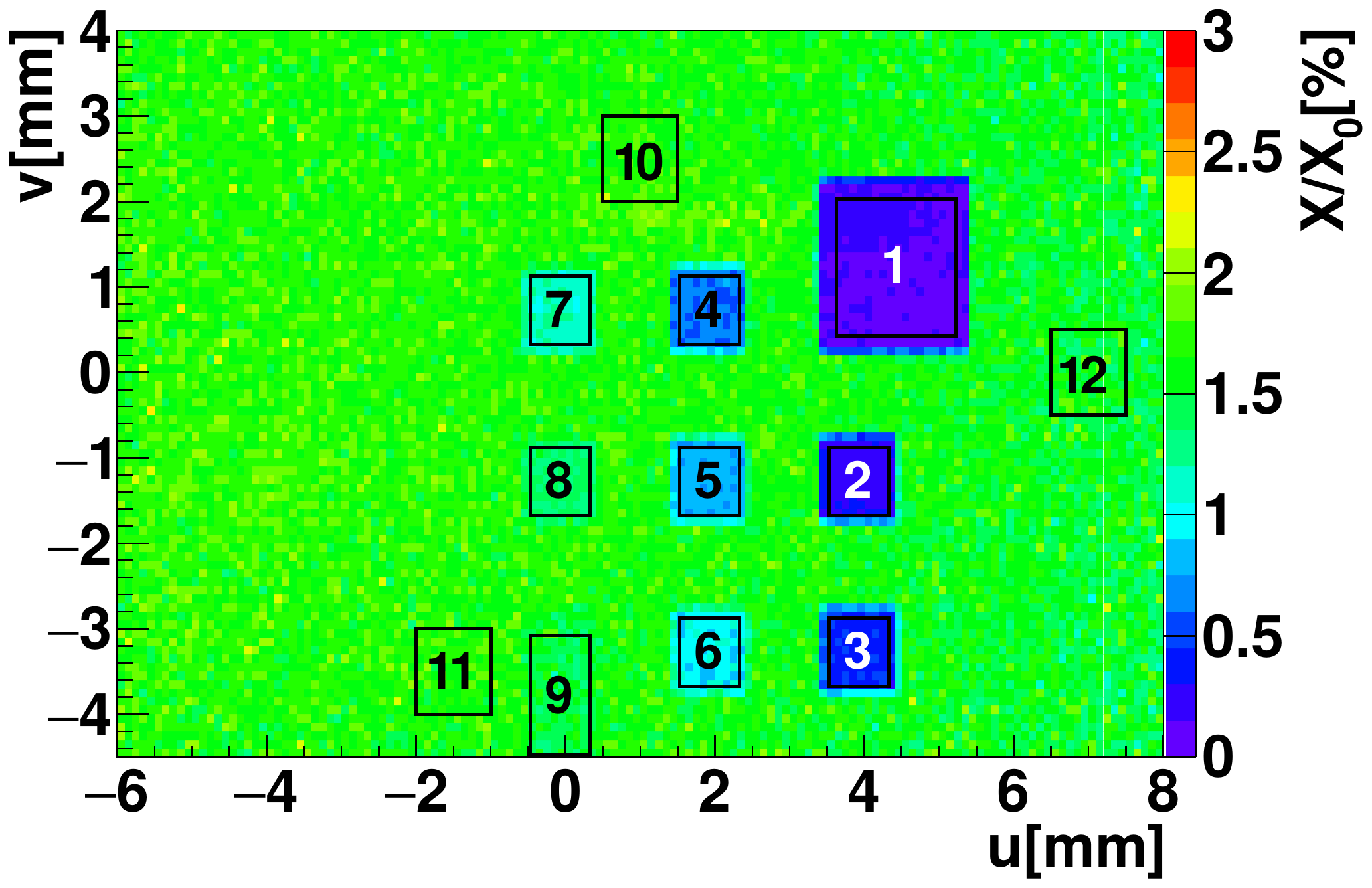}
  	\includegraphics[width= 0.60\linewidth]{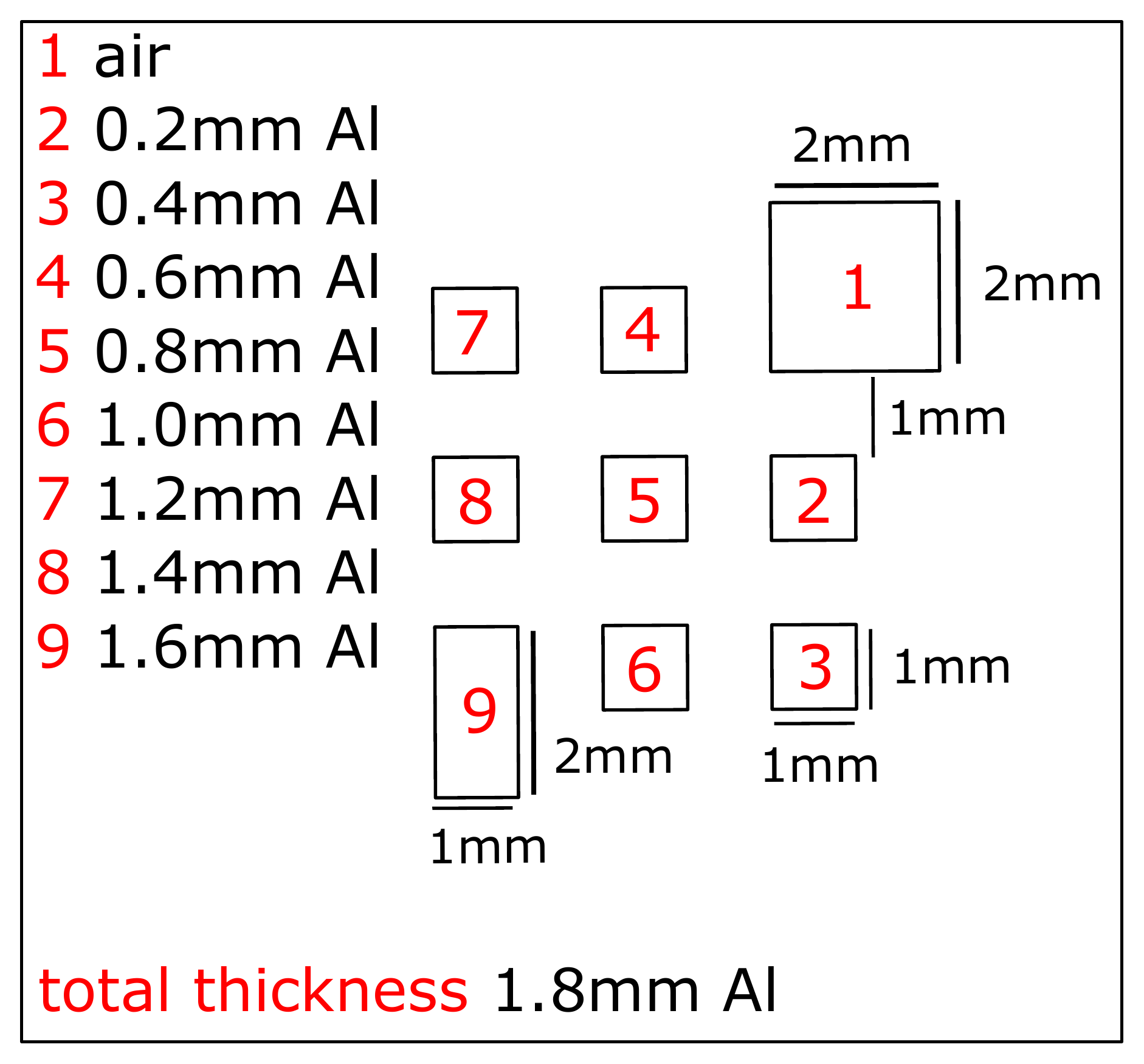}
 	\caption{{Uncalibrated ($\lambda=1.0$) radiation length image (upper panel) and schematic drawing (lower panel) of the aluminium calibration target. Tracks intersecting the 12 fiducial regions indicated by the boxes in the $X$/$X_0$ image are used for the calibration measurement.}}
 	 \label{fig:calibrationtarget}
\end{figure}

\begin{figure}[h!]
 	\centering
  	\includegraphics[width= 0.90\linewidth]{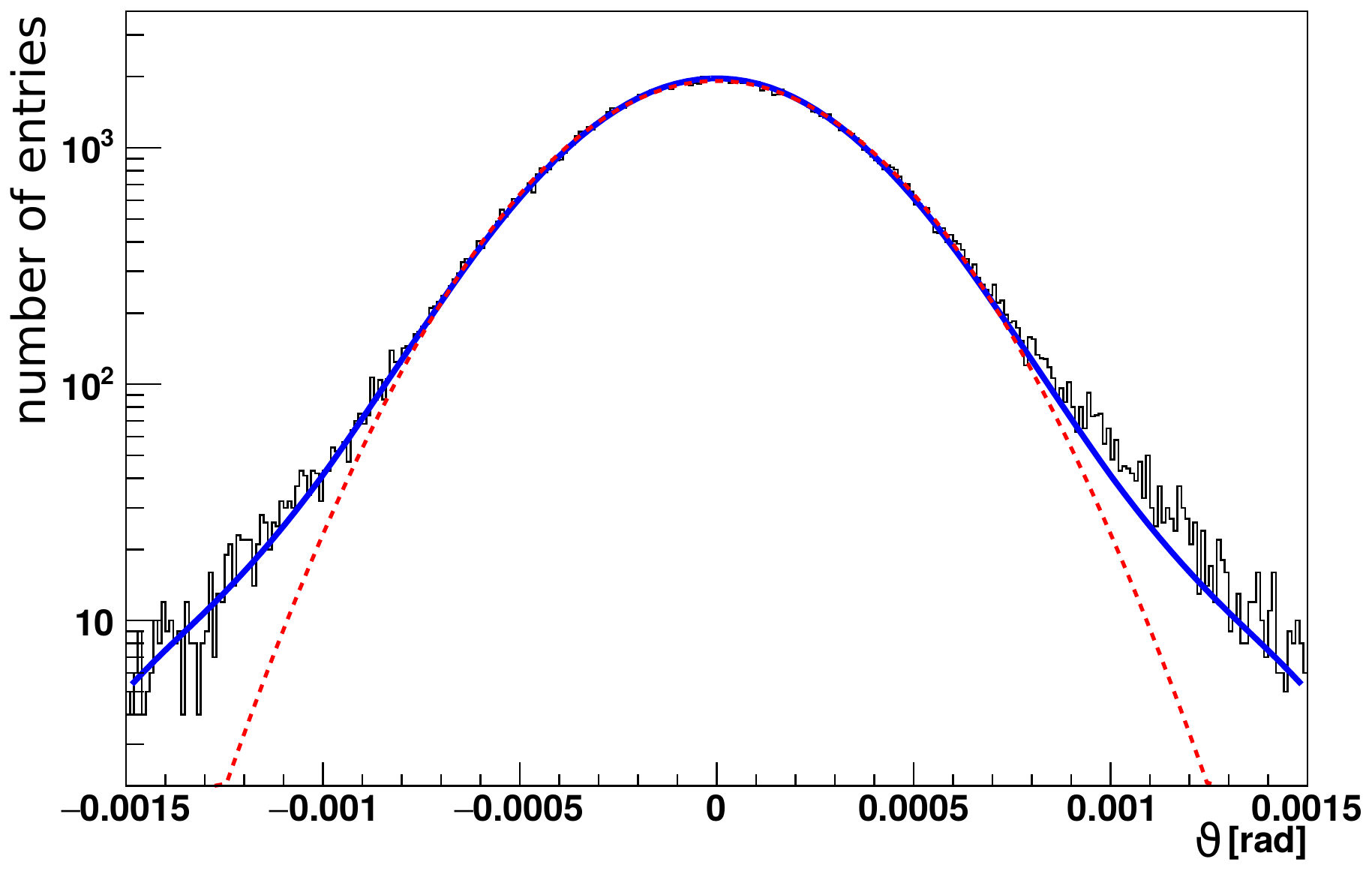}
 	\caption{{The reconstructed angle distribution (thin black line) and the fit function, based on the Moliere model (solid blue curve) and the Highland model (dashed red curve).}}
 	 \label{fig:molierefit}
\end{figure}

The image pixel size is mainly determined by statistics: Each pixel should have an angle distribution with roughly 1000 entries to ensure a stable fitting with the function in eq.\ref{eq:fitfunction}. The exact number of tracks per pixel is also dependent on the fraction between multiple scattering width $\sigma_\mathrm{HL}$ and angular resolution $\sigma_\mathrm{reso}$. For materials with a large radiation length fewer tracks per pixel can be used.\\
For precise $X$/$X_0$ measurements it is crucial to know the telescope angular resolution with a very small uncertainty. However, systematical effects like reference telescope alignment, approximations in the tracking model and the uncertainty on the M26 hit reconstruction resolution $\sigma_\mathrm{M26}$ are affecting the angular resolution in the measurement plane. The precise determination of the angular resolution with a calibration measurement is therefore vital.\\
For this purpose a calibrated aluminium target is employed. The calibration target (see lower panel in figure \ref{fig:calibrationtarget}) is a stack of 9 aluminium layers of $0.2\,$mm thickness with cutouts. The cutouts  define a 3 x 3 grid of fiducial regions with controlled material thicknesses reaching from air to $1.6\,$mm aluminium. Three additional fiducial regions have the full thickness of $1.8\,$mm. While the aluminium target is by far the thickest material in the setup, the energy loss for 4 GeV electrons is still small ($<\,$2$\,\%$) and therefore neglected here. The upper panel in figure \ref{fig:calibrationtarget} depicts the uncalibrated ($\lambda=1.0$) $X$/$X_0$ image with boxes indicating the 12 regions. During the calibration $\lambda$ is determined by a simultaneous $\chi^2$ fit of the reconstructed angle distributions in all 12 fiducial regions. For each fit function (eq. \ref{eq:fitfunction}) the thickness is fixed to the reference value. Figure \ref{fig:molierefit} depicts the reconstructed angle distribution of one of the fiducial regions and the corresponding fit based on the Moliere model (blue curve) and Highland model (red dashed curve). The core and the non-gaussian tail of the reconstructed angle distribution are modeled accurately by the Moliere model. The Highland model can be used to describe the core region of the distribution. For the radiation length measurement the core of the distribution is the most sensitive part. Because of this and the faster fitting procedure the Highland model was used in the following measurements.

 For the setup depicted in figure \ref{fig:Setup_PXD} the Highland fit yields a $\lambda$ of $1.152\pm0.003$. The corresponding $\lambda$ for the Moliere model is $1.165\pm0.003$. The small tension between the two results shows, that for consistent measurements a single multiple scattering model should be used. The composition of systematical effects causing the 15$\%$ deviation between expected and real angular resolution is not clear, but small contributions from different sources are expected. Among them are for example telescope misalignment and the uncertainty of the M26 hit position resolution $\sigma_\mathrm{M26}$.

\section{Results}
\label{sec:results}
\begin{figure}[h!]
 	\centering
  	\includegraphics[width= 0.940\linewidth]{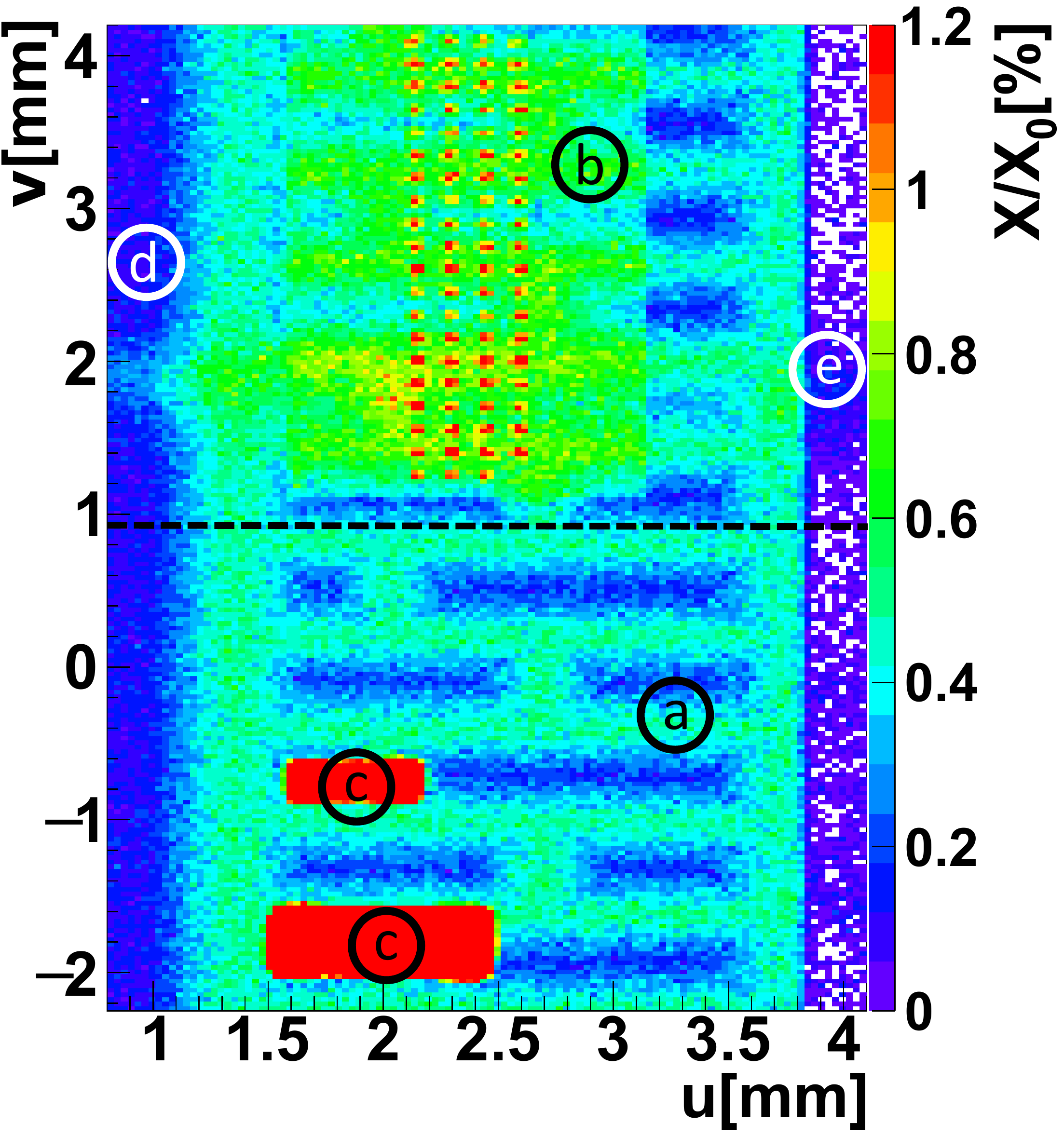}
	\includegraphics[width= 0.700\linewidth]{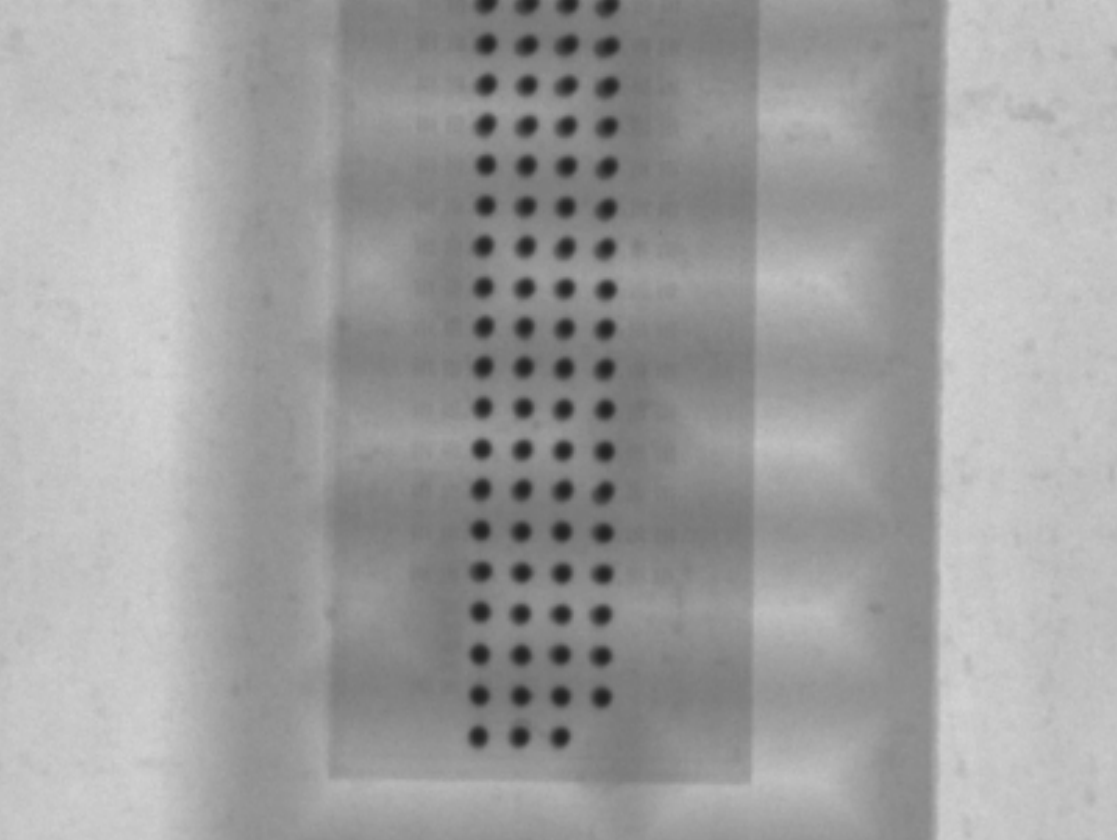}
 	\caption{{$X$/$X_0$ image (upper panel) and X-ray image (lower panel) of a mechanical prototype of a pixel module for Belle II. The labeled regions in the $X$/$X_0$ image are the thick silicon balcony with groove etchings (a), an ASIC (b), capacities (c), a $75\,\mu$m thin silicon membrane (d) and air (e). The X-ray image has a smaller field of view and shows the upper part of the $X$/$X_0$ image.}}
 	 \label{fig:pxdimage}
\end{figure}

To demonstrate the practical application of the $X$/$X_0$ imaging method we conducted a measurement on a mechanical Belle II pixel detector (PXD) and a silicon strip vertex detector (SVD) prototype module \cite{Abe:2010gxa}. The $X$/$X_0$ image of the mechanical PXD prototype can be seen in the upper panel of figure \ref{fig:pxdimage}.
The image area shows the thinned sensitive area and part of the balcony with the readout ASICs. It is based on a track sample of approximately 100 million tracks, which were covering an area of roughly 20 x $10\,\mathrm{mm}^2$. The pixel size is 30 x $30\,\mu\mathrm{m}^2$, so that a mean number of 1000 tracks per pixel is achieved. The bump bonds below the chip are clearly visible in the $X$/$X_0$ image. As a confirmation of the measured bump bond pattern a X-ray of a similar ASIC is depicted in the lower panel of figure \ref{fig:pxdimage}. \\
There are at least three areas in the image, where the high spatial resolution of the map can be validated. First of all the edge of the ASIC is clearly visible. The transition between silicon balcony and air can also be seen in the $X$/$X_0$  image as a sharp edge. The transition takes place within the borders of only one pixel. Another test of the spatial resolution is possible in the region of the solder bump bonds. The solder balls have a diameter of approximately $100\,\mu$m and the distance between two bumps is $150\,\mu$m \cite{Abe:2010gxa}. In the image this gives us a very periodic pattern of three pixels (corresponding to $90\,\mu$m) in between the bump bonds and two pixels ($\hat{=}\,60\,\mu$m) with an increased radiation length in the solder ball area.
\begin{figure}[h!]
 	\centering
  	\includegraphics[width= 0.990\linewidth]{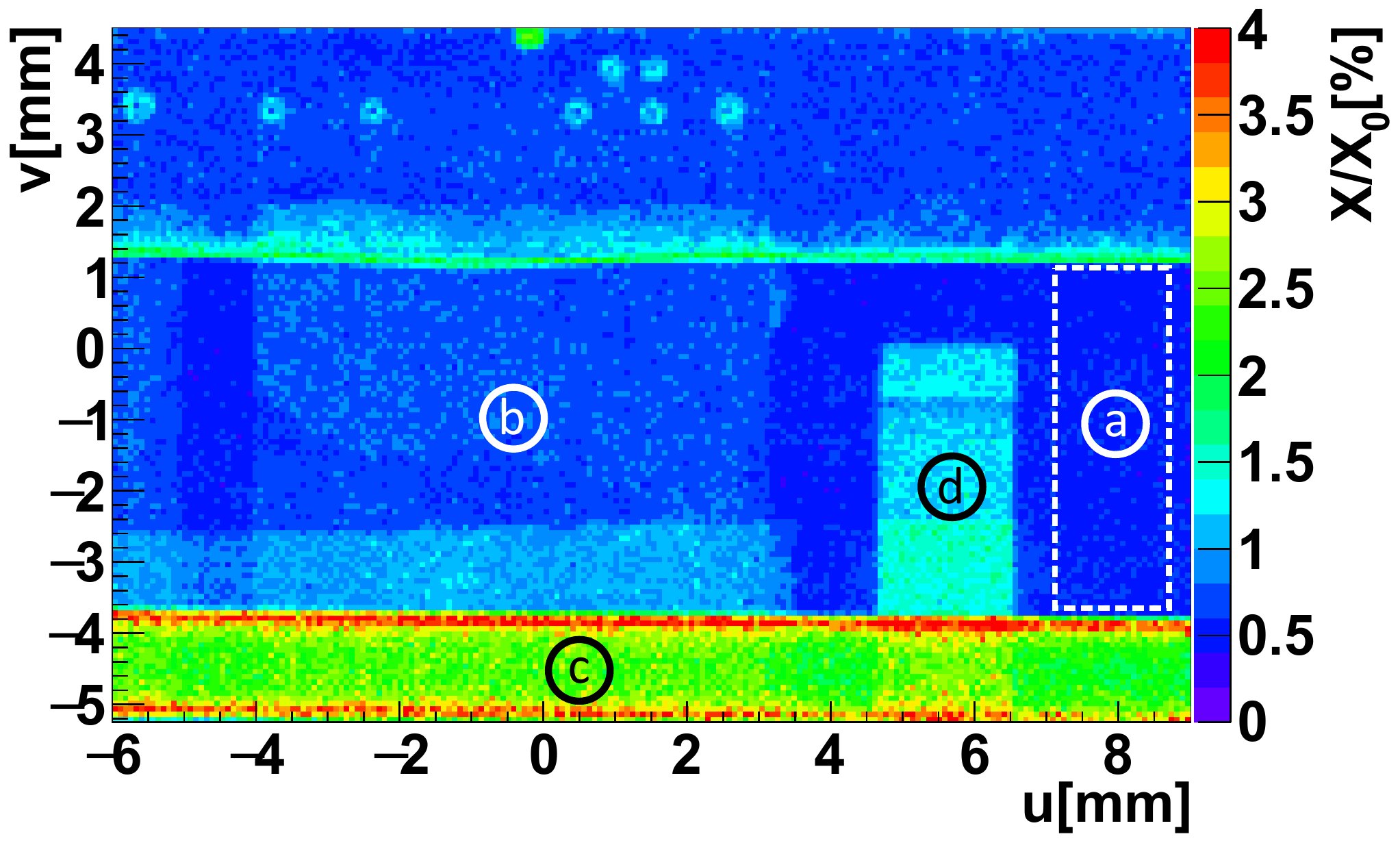}
	\includegraphics[width= 0.800\linewidth]{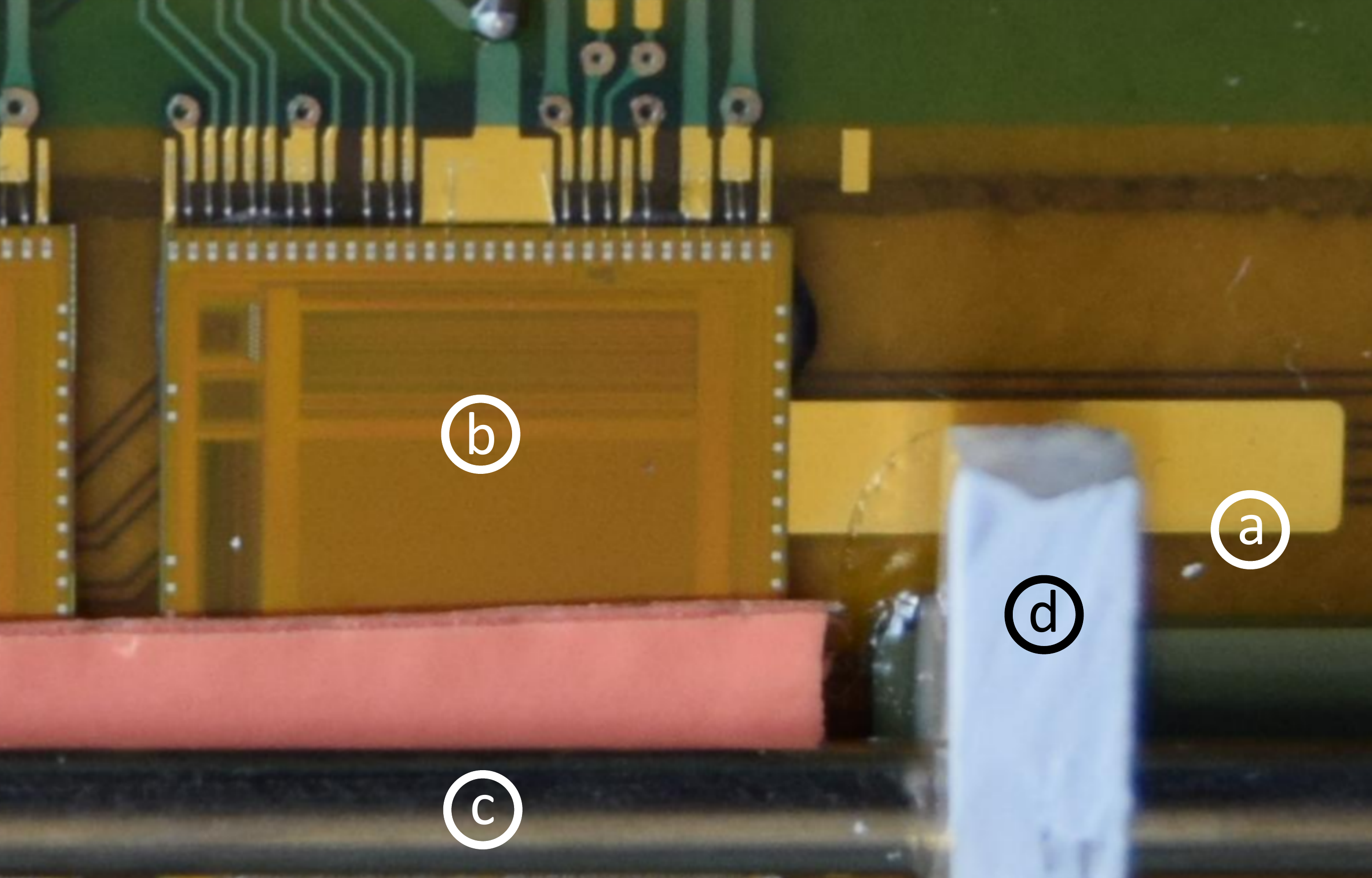}
 	\caption{{$X$/$X_0$ image (upper panel) and photograph (lower panel) of a SVD prototype. Different components can be seen: silicon sensor and polymide flex (a), ASICs (b), stainless steel cooling pipe (c) and a plastic clamp to support the cooling pipe (d). The horizontal line near the ASIC edge is due to a support structure on the back side of the module.}}
 	 \label{fig:svdimage}
\end{figure}
A second measurement was conducted with the mechanical prototype of the Belle II SVD module. The radiation length image compared to a photograph is shown in figure \ref{fig:svdimage}. A track sample of approximately 30 million tracks was used and the pixel size of the image is 75 x $75\,\mu\mathrm{m}^2$. The SVD is a very complex and highly integrated strip sensor, that consists of stacks of silicon, polymide flex and copper layers for routing. Additionally radiation length contributions from ASICs, plastic support structures and a cooling pipe can be seen in the image. \\
In order to demonstrate the precision of the radiation length measurements we choose a measurement area with a uniformly low material budget. In the area within the white dashed line (see figure \ref{fig:svdimage}) we find:
\begin{eqnarray}
\label{eq:origamiandsilicon}
X\mathrm{/}X_0= 0.544\pm0.002\,\left(\mathrm{stat.}\right)\pm 0.003\,\left(\mathrm{syst.}\right)\%
\end{eqnarray}
The statistical error can be computed from the fit errors of the single pixels, and the systematical error is computed from the error propagation of the statistical error of the calibration factor $\lambda$. In this region a 320 $\mu$m thick layer of silicon, an isolation layer and polymide flex is present, which also includes to 3 layers of $5\,\mu$m thick copper. The overall expected radiation length is approximately $0.5\,\%$ \cite{Abe:2010gxa}. The absolute difference between the radiation length values of the measurement and expectation is therefore well below $0.1\,\%$.

\section{Conclusion}
\label{sec:conclusion}

The paper presents a method to measure and image the radiation length of thin planar objects. The good spatial resolution of this imaging method can be used for example to measure the radiation length contribution of different components on a vertex detector module. Precise knowledge of the material distribution in the vertex detector layers is an important prerequisite for an an accurate vertex reconstruction and determination of track parameters.\\
In order to conduct the radiation length measurements three requirements must be covered. Firstly, a high-resolution telescope with an appropriate geometrical setup is needed to ensure a good angular and spatial resolution for the angle reconstruction. Secondly, a calibration measurement of the angular resolution must be conducted. This calibration step is crucial, because miscalibration of the angular resolution enters the radiation length estimation as an important systematical uncertainty.
The third requirement is a multi GeV particle beam to produce a sufficiently large multiple scattering signal. Additionally the spatial resolution of the image depends on the number of tracks per area. Consequently a beam with a large particle flux is needed to complete the measurement within a reasonable time frame. For a large enough track sample of 100 million tracks an image resolution of $30\,\mu$m can be achieved, which was validated by resolving bump bonds below ASICs.\\
\newline
The measurements leading to these results have been performed at the Test Beam Facility at DESY Hamburg (Germany), a member of the Helmholtz Association (HGF). This work was supported by the German Ministerium f\"ur Bildung, Wissenschaft, Forschung und Technologie (BMBF) and the VolkswagenStiftung.

\section*{References}


\end{document}